\documentclass[usenatbib]{mn2e}

\usepackage[dvips]{graphicx}
\usepackage{amssymb}
\usepackage{txfonts}

\newbox\grsign \setbox\grsign=\hbox{$>$} \newdimen\grdimen \grdimen=\ht\grsign
\newbox\simlessbox \newbox\simgreatbox \newbox\simpropbox
\setbox\simgreatbox=\hbox{\raise.5ex\hbox{$>$}\llap
     {\lower.5ex\hbox{$\sim$}}}\ht1=\grdimen\dp1=0pt
\setbox\simlessbox=\hbox{\raise.5ex\hbox{$<$}\llap
     {\lower.5ex\hbox{$\sim$}}}\ht2=\grdimen\dp2=0pt
\setbox\simpropbox=\hbox{\raise.5ex\hbox{$\propto$}\llap
     {\lower.5ex\hbox{$\sim$}}}\ht2=\grdimen\dp2=0pt
\def\ga{\mathrel{\copy\simgreatbox}}
\def\la{\mathrel{\copy\simlessbox}}
\def\simprop{\mathrel{\copy\simpropbox}}

\title[Core shifts of extragalactic jets]{Core shifts, magnetic fields and magnetization of extragalactic jets}

\author[A. A. Zdziarski et al.]
{Andrzej A. Zdziarski,$^1$ Marek Sikora,$^1$ Patryk Pjanka$^{2,1}$ and Alexander Tchekhovskoy$^{3,4}$\\
$^1$Centrum Astronomiczne im.\ M. Kopernika, Bartycka 18, PL-00-716 Warszawa, Poland\\
$^2$Obserwatorium Astronomiczne Uniwersytetu Warszawskiego, Al.\ Ujazdowskie 4, PL-00-478 Warszawa, Poland\\
$^3$Department of Physics and Department of Astronomy, University of California, Berkeley, CA 94720-3411, USA\\
$^4$Lawrence Berkeley National Laboratory, 1 Cyclotron Rd, Berkeley, CA 94720, USA
}

\date{Accepted 2015 May 1.  Received 2015 May 1; in original form 2014 October 16}

\pagerange{\pageref{firstpage}--\pageref{lastpage}}
\pubyear{2015}

\begin{document}

\maketitle

\label{firstpage}

\begin{abstract}
We study the effect of radio-jet core shift, which is a dependence of the position of the jet radio core on the observational frequency. We derive a new method of measuring the jet magnetic field based on both the value of the shift and the observed radio flux, which complements the standard method that assumes equipartition. Using both methods, we re-analyse the blazar sample of Zamaninasab et al. We find that equipartition is satisfied only if the jet opening angle in the radio core region is close to the values found observationally, $\simeq$0.1--0.2 divided by the bulk Lorentz factor, $\Gamma_{\rm j}$. Larger values, e.g., $1/\Gamma_{\rm j}$, would imply magnetic fields much above equipartition. A small jet opening angle implies in turn the magnetization parameter of $\ll 1$. We determine the jet magnetic flux taking into account this effect. We find that the transverse-averaged jet magnetic flux is fully compatible with the model of jet formation due to BH spin energy extraction and the accretion being a magnetically arrested disc (MAD). We calculate the jet average mass-flow rate corresponding to this model and find it consists of a substantial fraction of the mass accretion rate. This suggests the jet composition with a large fraction of baryons. We also calculate the average jet power, and find it moderately exceeds the accretion power, $\dot M c^2$, reflecting BH spin energy extraction. We find our results for radio galaxies at low Eddington ratios are compatible with MADs but require a low radiative efficiency, as predicted by standard accretion models.
\end{abstract}
\begin{keywords}
acceleration of particles--galaxies: jets--ISM: jets and outflows--magnetic fields--quasars: general--radiation mechanisms: non-thermal.
\end{keywords}

\section{Introduction}
\label{intro}

We study here extended jets, whose low-frequency emission originates in both a part of the jet which is optically thick to synchrotron self-absorption, and a part which is optically thin (\citealt{bk79}, hereafter BK79; \citealt{konigl81}). Then, the partially self-absorbed emission peaks at a distance $\simprop \nu^{-1}$ along the jet (where $\nu$ is the observed frequency), forming the radio core. This dependence of core position on frequency of observation is called the core shift. In this work, we study this effect theoretically. In particular, we consider the dependence of the magnetic field strength derived from the core shift on the jet radio flux.

Then, we consider the samples of blazars and radio galaxies of \citet{zamaninasab14}, hereafter Z14. We apply to them our theoretical results, and study equipartition and the jet opening angles. We re-consider the application by Z14 of the model of jet formation from black-hole (BH) spin-energy extraction \citep{bz77} with the accretion being magnetically arrested (MAD, \citealt*{nia03,mtb12,tnm11,tm12,t15}). Such flows have dragged so much magnetic flux to the BH that the flux becomes dynamically important and obstructs the accretion, hence the name. 

\section{Core shift in the model of Blandford \& K{\"o}nigl (1979)}
\label{core}

\citet{lobanov98} and \citet{hirotani05} have used core shifts to derive formulae for the magnetic field in the jet frame, $B(h)$, at a given distance, $h$, along the jet. In these formulae, the information about the jet radiative flux was not used. Then, the derived value of $B$ depends on the unknown normalization of the electron distribution. In order to specify it, they assumed a degree of equipartition between the energy density of the magnetic field and electrons, i.e., the plasma $\beta$ parameter that we define below in equation (\ref{beta}). However, this parameter can be far away from unity. Here we derive a formula for $B(h)$ that uses the information about the flux and makes no assumption about the equipartition.

We first re-consider the result of \citet{lobanov98} and \citet{hirotani05}. We use the formulation of the model of BK79 of \citet*{zls12}\footnote{Since the work of ZLS12 primarily concerns jets in BH binaries, their expressions do not include dependences on the cosmological redshift. To include them, the right-hand sides of equations (18) and (21) in ZLS12 need to be multiplied by $(1+z)$, and that of equation (22), by $(1+z)^{7/2}$. The powers of $(1+z)$ in remaining formulae follow from that, and they are $-(p+4)/2$ in equation (23), $(3-p)/2$ in equation (24), and $(p-3)/2$ in equation (26). Also, the power of the Doppler factor in equation (26) has been misprinted during typesetting; it should be $-(p+3)/2$.} (hereafter ZLS12). The jet emits above some minimum distance, $h_0$, over its length. Its total synchrotron spectrum consists of a low-frequency part, in which the jet is optically thick to synchrotron self-absorption up to some distance and optically thin further out, and a high-frequency part, which is optically thin. The boundary between these two parts is called the break frequency, $\nu_{\rm t}$ (which we define in the observer frame), and its value depends on $h_0$. The partially optically-thick regime has the energy spectral index of 0, and thus $F_\nu=$ constant at $\nu\la \nu_{\rm t}$. The model of BK79 assumes conservation of the relativistic-electron flux and the toroidal magnetic energy flux in a jet for which both the bulk Lorentz factor, $\Gamma_{\rm j}$, and the (half) opening angle, $\Theta_{\rm j}$, are constant. This implies
\begin{equation}
K(h)=K_0 (h/h_0)^{-2},\quad B(h)=B_0(h/h_0)^{-1},\quad N(\gamma,h) = K(h)\gamma^{-p},
\label{z_dep}
\end{equation}
where $N$ is the electron distribution, $K$ is its normalization, $\gamma$ the Lorentz factor of electrons, $\gamma_{\rm min}<\gamma\leq \gamma_{\rm max}$, and $p$ the electron index. 

Often, we know neither $\nu_{\rm t}$ nor $h_0$. However, in the partially optically-thick regime, which we consider here, the emission at a given frequency is mostly emitted by a narrow range of distance, peaking at $z\propto \nu^{-1}$. Thus, the actual value of $h_0$ is of no importance for emission below the break, $\nu_{\rm t}$. Therefore, we can parameterize the jet using the dependences (\ref{z_dep}) down to an arbitrary position, which we take at the gravitational radius, $r_{\rm g}\equiv GM/c^2$ (where $M$ is the BH mass). We denote the values of $B$ and $K$ at $h=r_g$ as $B_{\rm g}$, $K_{\rm g}$, respectively. We stress that this does not imply any jet emission there and merely provides a convenient parameterization. Then, we have 
\begin{equation}
K(h)=K_{\rm g} (h/r_{\rm g})^{-2},\quad B(h)=B_{\rm g}(h/r_{\rm g})^{-1},\quad \nu_{\rm t}(h_0)=\nu_{\rm g}(h_0/r_{\rm g})^{-1},
\label{z_dep2}
\end{equation}
where $\nu_{\rm g}$ is the observed break frequency at $h_0=r_{\rm g}$.
 
\begin{figure}
\centerline{\includegraphics[width=7.5cm]{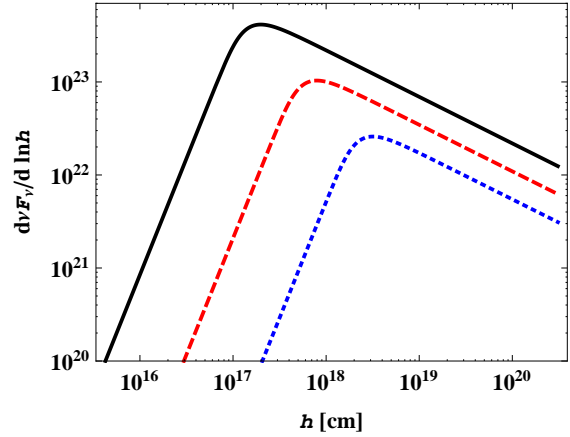}}
\caption{An example of the dependences of jet radio flux per unit $\ln h$ at three different frequencies, 2, 8 and 32 GHz shown by the blue dotted, red dashed and black solid curves, respectively, and at $z=0$. In this representation, the relative contribution of different distance ranges is proportional to the plotted curves. The peaks correspond to the position of the radio core. In this example, $p=2$ and the jet emits above $h_0\simeq 3\times 10^{13}$ cm. At this distance, the jet is optically thin at $\nu > \nu_{\rm g}\simeq 3\times 10^{14}$ Hz; the plotted curves are insensitive to this value.
}
\label{core_shift}
\end{figure}

The peak flux per unit $h$ is emitted at $h\simeq r_{\rm g}\nu_{\rm g}/\nu$. This follows, e.g., from equation (21) of ZLS12, which gives the jet observed flux following from the equation of radiative transfer, and which integrand gives the spatial profile of the emission. We can calculate those profiles in the optically thick and thin regimes, which give ${\rm d}F/{\rm d}h \propto h^{3/2}$ and $\propto h^{-(1+p)/2}$, respectively, see Fig.\ \ref{core_shift}, and the intersection at $h=(\upi/4)^{2/(p+4)} r_{\rm g}\nu_{\rm g}/\nu$, where the numerical factor is indeed close to unity for the usual $p>1$. Then, we can use the condition of the self-absorption optical depth at $r_{\rm g}$ being unity for $\nu_{\rm g}$, which can be obtained, e.g., from equation (23) of ZLS12,
\begin{equation}
\left[(1+z)h_{\rm P}\nu_{\rm g}\over m_{\rm e} c^2\right]^{p+4\over 2} ={\upi C_2(p) \sigma_{\rm T} K_{\rm g} r_{\rm g}\tan \Theta_{\rm j}\over \alpha_{\rm f}\sin i} \left(\delta B_{\rm g}\over B_{\rm cr}\right)^{p+2\over 2},
\label{et}
\end{equation}
where $h_{\rm P}$ is the Planck constant, $m_{\rm e}$ is the electron rest mass, $\sigma_{\rm T}$ is the Thomson cross section, $B_{\rm cr}={2\upi m_{\rm e}^2 c^3/(e h_{\rm P})}$ is the critical magnetic field, $\alpha_{\rm f}$ is the fine-structure constant, $C_2(2,\,3)= 2/3,\,1$, respectively, is a coefficient (for tangled $B$) defined in ZLS12, $z$ is the redshift, $\delta= [\Gamma_{\rm j}(1-\beta_{\rm j}\cos i)]^{-1}$ is the Doppler factor, $i$ is the viewing angle, and $\beta_{\rm j}c$ is the jet bulk velocity. (We follow here the notation of ZLS12 except that we denote the Doppler factor by $\delta$, and the distance along the jet by $h$.)

The core shift between two frequencies along the jet is,
\begin{equation}
\Delta h={D_L\Delta \theta\over (1+z)^2\sin i}= r_{\rm g}\nu_{\rm g}(\nu_1^{-1}-\nu_2^{-1}),
\label{d_z}
\end{equation}
where $\Delta \theta$ is the observed angular shift and $D_L$ is the luminosity distance. Here we can substitute $\nu_{\rm g}$ of equation (\ref{et}) to get $\Delta h$ in terms of $K_{\rm g}$ and $B_{\rm g}$. On the other hand, the quantity defined by \citet{lobanov98} and often reported in literature (e.g., in \citealt{pushkarev12} and Z14) for the core shift is 
\begin{equation}
\Omega_{r\nu}\equiv 1\,{\rm mas}{D_L[{\rm pc}]\Delta \theta[{\rm mas}] \over (\nu_1[{\rm GHz}]^{-1}-\nu_2[{\rm GHz}]^{-1})(1+z)^2},
\label{omegarv}
\end{equation}
which is the shift in pc per unit $1/\nu$ difference in GHz$^{-1}$. Since in our new method of measuring $B$, equation (\ref{b2}) below, we use separately $D_L$ and $\Delta \theta/ (\nu_1^{-1}-\nu_2^{-1})$, we can calculate the latter using $D_L$ calculated with the same cosmological parameters as those used by the original authors. Since we analyse below the sample of Z14, who used $\Omega_{r\nu}$ from \citet{pushkarev12}, we use the same parameters as them, $\Omega_\Lambda=0.73$, $\Omega_{\rm m}=0.27$ and $H_0=71$ km/(s Mpc). 

\citet{lobanov98} and \citet{hirotani05} assumed a degree of equipartition between the relativistic electrons and magnetic field. To quantify deviations from it, we use a convenient, but non-standard, definition of the plasma $\beta$ parameter, in terms of energy densities rather than pressures, 
\begin{equation}
\beta\equiv {u_{\rm p}\over B^2/8\upi}={K m_{\rm e} c^2 (1+k)f\over B^2/8\upi},\quad f=\cases{{\gamma_{\rm max}^{2-p}-\gamma_{\rm min}^{2-p}\over 2-p},& $p\neq 2$;\cr
\ln {\gamma_{\rm max}\over \gamma_{\rm min}},& $p=2$,}
\label{beta}
\end{equation}
where $u_{\rm p}$ is the particle energy density and $k$ takes into account the energy density in particles other than the power-law electrons, in particular in ions (excluding the rest energy). We use such a definition because the magnetic pressure depends on the field configuration, being $B^2/8\upi$ and $B^2/24\upi$ and for a toroidal field and fully tangled field, respectively, e.g., \citet{leahy91}, as well as the particle pressure depends on its adiabatic index, while the energy densities do not depend on those. Furthermore, the energy density ratio is customarily used for defining equipartition in astrophysics. 

From the above definition and $K/B^2=K_{\rm g}/B_{\rm g}^2$ (see equation \ref{z_dep2}), we obtain the magnetic field strength (in the comoving frame) at the distance $h$ (e.g., 1 pc) along the jet,
\begin{eqnarray}
\lefteqn{
B_\beta(h)=h^{-1}\times\nonumber}\\
\lefteqn{\qquad
\left[h_{\rm P} D_L\Delta \theta \over (\nu_1^{-1}-\nu_2^{-1})(1+z)\right]^{{p+4\over p+6}} \left(B_{\rm cr}\over \delta m_{\rm e}c^2 \sin i\right)^{{p+2 \over p+6}}\left[8 \alpha_{\rm f}(1+k)f \over \beta C_2(p) \sigma_{\rm T}  \tan\Theta_{\rm j}\right]^{{2 \over p+6}}
\nonumber}\\
\lefteqn{\qquad\quad\simeq
{1\over h}\left\{10^{-7.00} D_L[{\rm pc}]\Delta \theta[{\rm mas}] \over (\nu_1[{\rm GHz}]^{-1}-\nu_2[{\rm GHz}]^{-1})(1+z)\right]^{{p+4\over p+6}} \left(10^{19.73}\over \delta \sin i\right)^{{p+2 \over p+6}}\times
\nonumber}\\
\lefteqn{\quad\qquad
\left[10^{23.12}(1+k)f \over \beta \tan\Theta_{\rm j}\right]^{{2 \over p+6}},
\label{b1}}
\end{eqnarray}
where the last of the numerical coefficients has been calculated for $p=2$. This expression is almost the same as the corresponding equation (43) of \citet{hirotani05} except\footnote{\label{ftn:z} This difference appears due to the assumption of \citet{hirotani05} that the special-relativistic invariance includes $z$, see his equations (17--18), which leads to the transformation of $\sin i$ from the jet to the observer frame including $(1+z)$. However, the transformation from the jet frame to the stationary frame at the jet redshift is $\sin i= \delta^{-1}\sin i'$. Then, the photon travels to the observer radially without changing its direction, as well as the jet axis is no more a distinct direction. Consequently, the coefficient of $(1+z)^{-1}$ appears in the expressions for $p=2$ of \citet{og09}, \citet{pushkarev12} and Z14 (who used the result of \citealt{hirotani05}) instead of $(1+z)^{-3/4}$. Also, equation (10) of \citet{lobanov98}, giving $B$ as a function of the core shift, has an apparent typo in the power of $\sin i$, which should be $-1/2$ instead of $-5/4$.} for his dependence on $(1+z)$. The numerical coefficient at $p=2$, $k=z=0$, $\gamma_{\rm max}/\gamma_{\rm min}=10^{4.34}$ (corresponding to $f=10$ assumed by \citealt{hirotani05}), $z$ and $D_L$ in pc, $\nu$ in GHz and $\Delta\theta$ in mas is $1.42\times 10^{-8}$, which is a small difference with respect to $1.45\times 10^{-8}$ in the expression of \citet{og09} (including their coefficient for $\Omega_{r\nu}$), due to rounding errors in the latter.

On the other hand, the above method ignores the information contained in the flux in the partially self-absorbed spectrum, $F_\nu$. Under the assumptions of BK79, this flux can be used to derive the magnetic field without linking the normalization of the electron distribution to equipartition. The most direct way to do it appears to express the observed flux via the source function of non-thermal electrons (which is independent of $K$ because both the emission and absorption coefficients are proportional to it) and then integrate the standard radiative transfer solution over the projected area of the jet. This has been done in ZLS12 in their equations (17--18), which allowed them to get a general expression for the flux in their equation (21) (dependent on $K$ only via the optical depth). Here, we use equation (22) of ZLS12, which is the limit of their equation (21) in the optically-thick regime (in which case the flux is completely independent of $K$), to relate $F_\nu$ [erg/(cm$^2$ s Hz)] to the magnetic field strength and the break energy at $r_{\rm g}$, $B_{\rm g}$ and $h\nu_{\rm g}$, respectively, for which we use equations (\ref{z_dep2}) and (\ref{d_z}) above. This way, we derive
\begin{eqnarray}
\lefteqn{
B_F(h)=  \!
{D_L\delta(h_{\rm P}/ m_{\rm e})^7\over h\left[(1+z)\sin i\right]^3  c^{12}}\left(B_{\rm cr}\Delta \theta \over \nu_1^{-1}-\nu_2^{-1}\right)^5\! \left[\alpha_{\rm f} C_1(p)C_3(p)\tan\Theta_{\rm j} \over 24\upi^3 C_2(p)F_\nu \right]^2\nonumber}\\
\lefteqn{\qquad
\simeq {3.35\times 10^{-11} D_L[{\rm pc}]\,\delta\, \Delta \theta[{\rm mas}]^5\tan\Theta_{\rm j}^2\over h[{\rm pc}](\nu_1[{\rm GHz}]^{-1}-\nu_2[{\rm GHz}]^{-1})^5\left[(1+z)\sin i\right]^3 F_\nu[{\rm Jy}]^2},
\label{b2}}
\end{eqnarray}
where $C_1(2,\,3)\simeq 1.14,\,1$, $C_3(2,\,3)\simeq 3.61,\,2.10$, respectively, are coefficients defined in ZLS12, and the numerical coefficient in the second line has been calculated for $p=2$. The weak dependence on $p$, only via the $C$ constants, is a consequence of the analogous lack of a dependence of the flux on $p$ in the partially self-absorbed part of the jet spectrum, see, e.g., equation (22) of ZLS12. The flux, $F_\nu$, is that of an inner jet integrated over both its optically thick and thin parts, see Fig.\ \ref{core_shift}. It is neither the flux of a radio core alone nor the flux including outer parts, e.g., radio lobes. The used flux should be of the flat part of the spectrum with $F_\nu\simprop \nu^0$. The relatively high power of $\Delta\theta$ above requires its accurate measurement. Both formulae for $B(h)$ are independent of $M$. 

We can also calculate the equipartition coefficient using equations (\ref{b1}--\ref{b2}). We obtain
\begin{eqnarray}
\lefteqn{
\beta={f(1+k) m_{\rm e}^{20+3p}c^{34+5p}\sin^{8+p} i\over D_L\sigma_{\rm T}h_{\rm P}^{19+3p}\delta^{4+p}}\left(\Delta \theta \over \nu_1^{-1}-\nu_2^{-1}\right)^{-13-2p} \left[8(1+z)\over \tan\Theta_{\rm j} B_{\rm cr}^2\right]^{7+p}\times\nonumber}\\
\lefteqn{\qquad
\left(C_2\over \alpha_{\rm f}\right)^{5+p} \left(3\upi^3 F_\nu\over C_1 C_3\right)^{6+p},
\label{beq}}
\end{eqnarray}
which is independent of $h$. The numerical coefficient at $p=2$, $f=1$, $k=0$, $D_L$ in pc, $\nu$ in GHz, $F_\nu$ in Jy and $\Delta\theta$ in mas is $3.28\times 10^{9}$. Given the high powers of most of the measured quantities, an application of this formula can yield a relatively large fractional error.

We note that radio cores allow for one more method to determine their magnetic field if their angular size is known (e.g., \citealt{slish63,williams63,hirotani05}), independently of the value of the source distance. For example, equation (39) of \citet{zdz14} can be used. Note that the flux in that formula corresponds to the core only, unlike that in equation (\ref{b2}) above.

\section{Equipartition in the sample of Z14}
\label{z14}

We re-analyse the sample of Z14, which contains 68 blazars and 8 radio galaxies. This allows us to study in detail their average properties. Hereafter, we assume $\Theta_{\rm j}\ll 1$, implying $\tan\Theta_{\rm j}\simeq \Theta_{\rm j}$. In this Section, we discuss blazars only, and consider the radio galaxies separately in Section \ref{RG}. 

Z14 made some approximate substitutions for blazars that we also adopt. Namely, we assume $\Gamma_{\rm j}= (1+\beta_{\rm app}^2)^{1/2}$, which is the minimum Lorentz factor for a given observed apparent velocity, $\beta_{\rm app}c$, $i=\Gamma_{\rm j}^{-1}$ (in which case $\delta\simeq \Gamma_{\rm j}$), and $\Theta_{\rm j}=0.13\Gamma_{\rm j}^{-1}$ (from the fit of \citealt{pushkarev09})\footnote{These assumptions were used in obtaining equation (4) from equation (3) in Z14.}. Following Z14, we also choose $p=2$, $f=10$, $k=0$ and $\beta=1$, but we obtain slightly higher values of $B_\beta$ than Z14, owing to our formula (\ref{b1}) having the corrected power of $(1+z)$ (see footnote \ref{ftn:z}). 

To calculate the magnetic field, $B_F$, from equation (\ref{b2}), we need to know the flux. For the sample of blazars, we get $F_\nu$ measured at 15 GHz at the same time as the core shift from the MOJAVE \citep{lister09} web page\footnote{http://www.physics.purdue.edu/astro/MOJAVE/allsources.html}. In a few cases, the core shift was based on two observations and/or two radio bands, for which we took the average flux. We do not consider the upper limits on the core shift, of 9 blazars in the sample of Z14, which leaves 59 blazars in our sample.  

\begin{figure}
\centerline{\includegraphics[width=7.5cm]{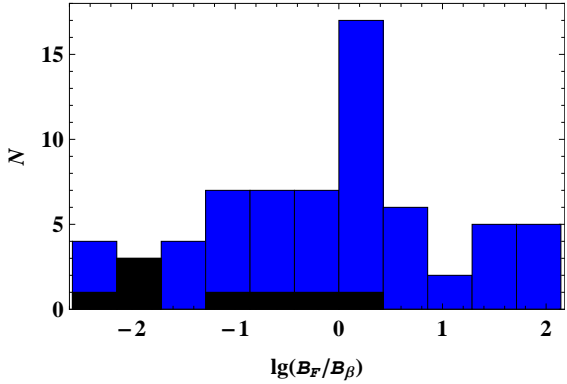}}
\caption{The histogram of the ratio of the magnetic field determined using the flux, equation (\ref{b2}), to that assuming pressure equipartition ($\beta=1$), equation (\ref{b1}), at $p=2$, $k=1$ and $\vartheta_{\rm j}= 0.13$. The blue and black colours indicate the blazars and radio galaxies, respectively.
}
\label{B_ratio}
\end{figure}

However, as Z14 note, the relationship between $\Theta_{\rm j}$ and $\Gamma_{\rm j}$ is observationally determined at larger distances than the radio cores, and it is not known at the radio cores. In fact, Z14 used $\Theta_{\rm j}=\Gamma_{\rm j}^{-1}$ in some of their following calculations. Thus, we allow the coefficient of the opening angle for blazars, $\vartheta_{\rm j}$, to be a free parameter, defined by $\Theta_{\rm j}\equiv \vartheta_{\rm j} /\Gamma_{\rm j}$.

Fig.\ \ref{B_ratio} shows the histogram of $B_F/B_\beta$ at $\vartheta_{\rm j}=0.13$ and $k=1$ for the 59 considered blazars, and at the values of $\vartheta_{\rm j}$ and $i$ listed in Table 1 for 8 radio galaxies. For blazars, we obtain $\langle B_F/B_\beta\rangle\simeq 1.6$ (hereafter the symbol $\langle\rangle$ denotes a geometric average, i.e., based on averaging the logarithms), and we see a pronounced peak in the distribution at $B_F/B_\beta\simeq 2$--3. This argues for a value of $\beta$ with a relatively small intrinsic dispersion in the blazar sample. We consider the errors in determining the core shift to be a major cause of the relatively large scatter seen in Fig.\ \ref{B_ratio}, corresponding to the standard deviation of $\lg B_F/B_\beta$ of $\simeq 1.1$. We have also checked that there is no statistically significant dependence of $B_F/B_\beta$ on $z$.

We define a value of $\vartheta_{\rm j}$ for blazars at which $\langle B_F/B_\beta\rangle=1$, which we denote by $\vartheta_0$. We then assume all sources have the same $\vartheta_{\rm j}$ and $\beta=k=1$. We find $\vartheta_0\simeq 0.11$ at $\beta=1$, and $\langle B_\beta\rangle \simeq 1.3$ G at 1 pc. On the other hand, we have the following dependences from equations (\ref{b1}--\ref{beq}),
\begin{equation}
\vartheta_{\rm j}/ \vartheta_0=\langle\beta\rangle^{-1/(p+7)}, \quad  \langle B_F/B_\beta\rangle=\langle\beta\rangle^{-2/(p+6)}=(\vartheta_{\rm j}/ \vartheta_0)^{2(p+7)/(p+6)}.
\label{B_beta}
\end{equation}
Thus, the second expression above implies, for $p=2$, that the distribution of blazars in Fig.\ \ref{B_ratio} can be shifted along the $B_F/B_\beta$ axis by $(\vartheta_{\rm j}/0.13)^{9/4}$, preserving the shape of the histogram. Then, we can have $\vartheta_{\rm j}>\vartheta_0$ if $\beta<1$. Still, $\langle B_F/B_\beta\rangle \simeq 160$ for the observed sample at $\vartheta_{\rm j}=1$. This corresponds to a very large departure from equipartition, $\langle \beta\rangle\simeq 1.6\times 10^{-9}$, with the standard deviation of $\lg \beta$ of $\simeq 4.4$. Since such large departures from equipartition are unlikely, this argues for $\langle\vartheta_{\rm j}\rangle\ll 1$ in the blazar sample (in agreement with the observations, see above). Hereafter, we assume $\vartheta_{\rm j}=\vartheta_0\simeq 0.11$ for blazars, corresponding to $\langle\beta \rangle=1$. On the other hand, $\beta$ moderately less than unity may occur if equipartition with magnetic field corresponds to only its random component, rather than the total one, being predominantly structured and toroidal.

We note that \citet{kovalev05} provides the angular sizes and fluxes of the cores at 15 GHz for all blazars in the studied sample. Thus, one could, in principle, obtain one more measurement of $B$ based on those data, see the last paragraph of Section \ref{core}. We leave it to a future study.

\section{Implications for the jet physics}
\label{physics}

The magnetization parameter is the ratio of the Poynting flux to the kinetic energy flux in the BH frame. In the jet rest frame, it is equal to the ratio of the proper magnetic enthalpy, $w_B$, to that for particles including the rest energy, $w_{\rm p}$,
\begin{equation}
\sigma = {w_B\over w_{\rm p}}\simeq {B^2/4\upi\over \eta u_{\rm p}+\rho c^2},
\label{magn}
\end{equation}
where $\rho$ is the rest-mass density, $4/3<\eta<5/3$ is the particle adiabatic index and $B=\langle B'_\phi\rangle$, i.e., it corresponds to the toroidal component of the jet-frame magnetic field. Here, we average over the jet cross section, i.e.,
\begin{equation}
B^2\equiv {2\upi \int_0^{r_{\rm j}}{\rm d}r\, r B'_\phi(r)^2\over \upi r_{\rm j}^2},
\label{avb}
\end{equation}
where $r_{\rm j}$ is the jet cylindrical radius, and we define $\rho$ and $u_{\rm p}$ as analogous averages. Since we measure the magnetic fields using synchrotron self-absorption (with the absorption coefficient $\propto B^{(2+p)/2}$ which is $B^2$ for $p=2$) through the entire source, the measured strength approximately corresponds to the above $B$ rather than, e.g., the surface value of $B'_\phi(r_{\rm j})$. Also, the jet magnetic power is strictly proportional to the above averaged value. 

Using equation (\ref{beta}), $1/\sigma=\eta \beta/2 + 4\upi\rho c^2/B^2$. If we know both $\sigma$ and $\beta$, we can constrain the plasma parameters,
\begin{equation}
{u_{\rm p}\over \rho c^2}={\beta\sigma/2\over 1-\beta\sigma\eta/2}, \quad {B^2/4\upi\over \rho c^2}={\sigma\over 1-\beta\sigma\eta/2}.
\label{u_e}
\end{equation}
We note that this implies $u_{\rm p}=2 \rho c^2$ at $\sigma=\beta=1$ and $\eta=3/2$ (corresponding to a mixture of relativistic electrons and nonrelativistic protons). This puts strong constraints on the electron distribution, requiring it to be hard in the absence of e$^\pm$ pairs. On the other hand, there is no such problems at $\sigma\ll 1$, which we find here. 

Indeed, $\sigma$ is approximately related to the opening angle,
\begin{equation}
\Theta_{\rm j}\simeq s \sigma^{1/2}/\Gamma_{\rm j},\quad \sigma\simeq (\vartheta_{\rm j}/ s)^2,
\label{open_sigma}
\end{equation}
where $s\la 1$ \citep*{tmn09,komissarov09}. Since $\langle\vartheta_{\rm j}\rangle\sim 0.1$--0.2 at least at large distances beyond the radio core \citep{pushkarev09,clausen13,jorstad05}, the average value of $\sigma$ there is $\ll 1$. A similar situation occurs in BH binaries, whose jets have the opening angles $\ll 1/\Gamma_{\rm j}$ \citep*{millerjones06}. Thus, some mechanisms in the jet have to be able to decrease $\sigma$ from an initial value of $\sim\Gamma_{\rm max}\gg 1$ to below unity, see \citet{komissarov11}, \citet{tmn09}, \citet{lyubarsky10}, where $\Gamma_{\rm max}$ is the Lorentz factor corresponding to the conversion off all of the magnetic energy into acceleration. Hereafter, we define $\Theta_{\rm j}$ as the ratio of the jet radius to the distance from the BH centre. 

The Bernoulli equation for negligible energy losses is,
\begin{equation}
\Gamma_{\rm j}\left[1+{\eta u_{\rm p}+B^2/4\upi\over \rho c^2}\right]= \Gamma_{\rm j}{1+\sigma\over 1-\beta\sigma\eta/2} =\Gamma_{\rm max}.
\label{bernoulli}
\end{equation}
Since this equation is obtained by dividing the equations of energy and mass conservation, the quantities involved are again averages over the jet cross section. This equation gives $\Gamma_{\rm j}$ as a function of $\sigma$ and $\beta$. We can then use equations (\ref{u_e}--\ref{bernoulli}) together with the conservation of mass, equation (\ref{mdot}) below, to determine the evolution of the toroidal magnetic field and particle energy density along the jet. We derive
\begin{equation}
B^2(h) =B_0^2 \beta_{\rm j}{h_0^2\over h^2}{1+\sigma_0\over 1+\sigma},\quad u_{\rm p}=u_{\rm p0}\beta_{\rm j}{h_0^2\over h^2}{\beta(1+\sigma_0)\over 1+\sigma},
\label{B_sigma}
\end{equation}
where $B(h)$ differs from the corresponding dependence of equation (\ref{z_dep}), which assumes constant $\Gamma_{\rm j}$ and $\Theta_{\rm j}$, whereas they vary in the present approach. Here $B_0$ and $u_{\rm p0}$ are the magnetic field strength and the particle energy density, respectively, at some reference point. In the evolution of $u_{\rm p}$, $\beta$ is likely to vary as well. Note that equation (\ref{B_sigma}) also conserves the enthalpy flux in the limit $\Gamma_{\rm j}\gg 1$, see equation (\ref{P_tot}) below. If $\Gamma_{\rm j}\sim 1$, e.g., near the jet base, both equation (\ref{z_dep}) and the relationship $\sigma\simeq (\Gamma_{\rm j}\Theta_{\rm j}/s)^2$ no longer hold. 

We then consider conservation of the poloidal magnetic flux in the model with extraction of the rotational power of the BH \citep{bz77}. The poloidal component of the magnetic field is non-uniform across the jet as a result of the conversion of the Poynting flux into the jet kinetic energy and it can be approximated as a power law in the radial coordinate with the index of $\alpha(\sigma)\sim 1$ \citep{tmn09}, $B_{\rm pj}(r/ r_{\rm j})^{-\alpha}$, where $B_{\rm pj}$ is the surface value (depending on $h$). The flux is approximately given by
\begin{equation}
\Phi_{\rm j}\equiv 2\upi \int_0^{r_{\rm j}}{\rm d}r\, r B_{\rm p}(r)={2 \over 2-\alpha}\upi B_{\rm pj}r_{\rm j}^2. 
\label{poloidal}
\end{equation}
The factor of $2/(2-\alpha)$ represents the correction for the non-uniformity, which was shown by \citet{tmn09} to be equal to $(1-\Gamma_{\rm j}/\Gamma_{\rm max})^{-1}\simeq (1+\sigma)/\sigma$ (where the right-hand side is, in the considered cold MHD approximation, for $\beta=0$). Thus,
\begin{equation}
\alpha={2\over 1+\sigma},\quad \sigma={2-\alpha\over \alpha}.
\label{alpha}
\end{equation} 
Then, the toroidal field component in the jet frame is given by
\begin{equation}
B'_\phi(r) = {\Omega_{\rm f} r\over c\Gamma_{\rm j}}B_{\rm pj}\left(r\over  r_{\rm j}\right)^{-\alpha}={\Omega_{\rm f} r_{\rm j}\over c\Gamma_{\rm j}}B_{\rm pj}\left(r\over  r_{\rm j}\right)^{1-\alpha}\equiv B'_{\phi{\rm j}}\left(r\over  r_{\rm j}\right)^{1-\alpha},
\label{bphi}
\end{equation}
where $\Omega_{\rm f}$ is the angular frequency of the field lines. For comparison with observations, we argued above that the average value of $B'_\phi$, see equation (\ref{avb}), should be used, which is
\begin{equation}
B=B'_{\phi{\rm j}}\left(1+\sigma\over 2\sigma\right)^{1/2}.
\label{bphiav}
\end{equation}

We can then write the magnetic flux in terms of the transverse-average magnetic field strength, $B$, as
\begin{equation}
\Phi_{\rm j}={2^{3/2}\upi r_{\rm H}s h B(1+\sigma)^{1/2}\over \ell a},
\label{phi_jet}
\end{equation}
where $r_{\rm H}=[1+(1-a^2)^{1/2}]r_{\rm g}$ is the BH horizon radius, $a$ is the dimensionless spin parameter, and $\ell\la 0.5$ is the ratio of $\Omega_{\rm f}$ to the BH angular frequency. The above equation is modified with respect to equation (5) of Z14, who expressed it in terms of $B'_{\phi{\rm j}}$ rather than the observed (average) field strength, and did not explicitly consider the relationship between the jet opening angle and $\sigma$, equation (\ref{open_sigma}). The above formula is identical to that of Z14 for $\sigma= 1$ and lower by $2^{-1/2}$ for $\sigma\ll 1$. In numerical calculations, $a=1$ is assumed, as in Z14. We determine the values of $B$ using equation (\ref{b1}), i.e., $B=B_\beta$, but impose the condition of $\langle B_F\rangle=\langle B_\beta\rangle$, which implies $\vartheta_{\rm j}\simeq 0.11$ for blazars, see Section \ref{z14}. We do not directly use the values of $B_F$ due to their large errors associated with the high power of the core shift, see Section \ref{core}.

We can equate $\Phi_{\rm j}$ to the poloidal flux treading the BH on one hemisphere, $\Phi_{\rm BH}$, which is limited by the ram pressure of the accretion flow \citep{nia03}. This can be written as
\begin{equation}
\Phi_{\rm BH}=\phi_{\rm BH}(\dot M c)^{1/2} r_{\rm g},
\label{phi_BH}
\end{equation}
where saturation values of $\phi_{\rm BH}\simeq 50$ have been found in GRMHD simulations of magnetically arrested accretion (\citealt{tnm11}; \citealt{mtb12}). Z14 estimated $\dot M$ as $L/\epsilon$, where $L$ is the estimated bolometric luminosity, assuming the radiative efficiency of $\epsilon=0.4$. However, the efficiency in the MAD scenario is not determined by the bounding energy on the marginally stable orbit of the standard accretion disc. Since the accretion proceeds via the interchange instability, it is very difficult to determine its efficiency. According to \citet{nia03}, such an efficiency can be $\sim$0.5, even for the Schwarzschild geometry. However, noting that a fraction of gravitational energy of accreting matter can be transmitted outwards mechanically via magnetic fields, lacking information what this fraction is we are not in position to determine radiative efficiency of the MAD. Here we assume $\epsilon=0.2$. For the blazar sample of Z14, $\ell=0.5$, $a=s=k=\beta=1$, and $\eta=3/2$, we obtain $\langle\phi_{\rm BH}\rangle\simeq 33(\epsilon/0.2)^{1/2}$, with the standard deviation of $\lg \phi_{\rm BH}$ of 0.23. Within the uncertainties of $\epsilon$, $\vartheta_{\rm j}$, $\beta$, $a$, and systematic uncertainties of our adopted idealized model, this is in full agreement with the simulation results. Thus, our results confirm that blazars can have jets originating from magnetically arrested discs and be powered by the Blandford-Znajek mechanism, but at the same time they can have $\sigma\ll 1$ at the radio core. 

\begin{figure}
\centerline{\includegraphics[width=8cm]{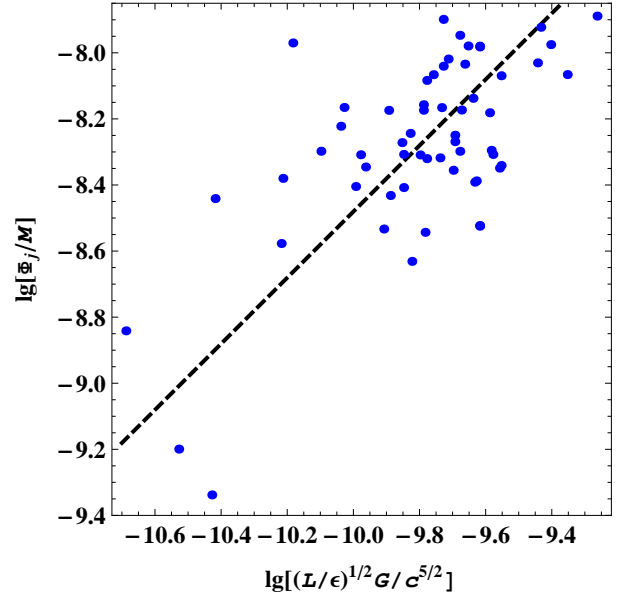}}
\caption{The correlation between $\Phi_{\rm j}/M$ and $L^{1/2}$ (normalized as in equation \ref{phi_BH}) for blazars. The dashed line corresponds to $\langle\phi_{\rm BH}\rangle$. 
}
\label{no_mass}
\end{figure}

Note that both $\Phi_{\rm j}$ and $\Phi_{\rm BH}$ depend linearly on the BH mass. Thus, the actual strength of the correlation is completely independent of it. Therefore, we present the correlation (for blazars only) for both quantities divided by $M$ in Fig.\ \ref{no_mass}. We see a relatively good correlation for blazars, though the visual scatter is much larger than in fig.\ 2 of Z14, where the common dependence on $M$ was included. This is also because we do not show here radio galaxies, discussed below in Section \ref{RG}.

\section{The jet mass-flow rate, composition, and power}
\label{power}

\begin{figure}
\centerline{\includegraphics[width=8cm]{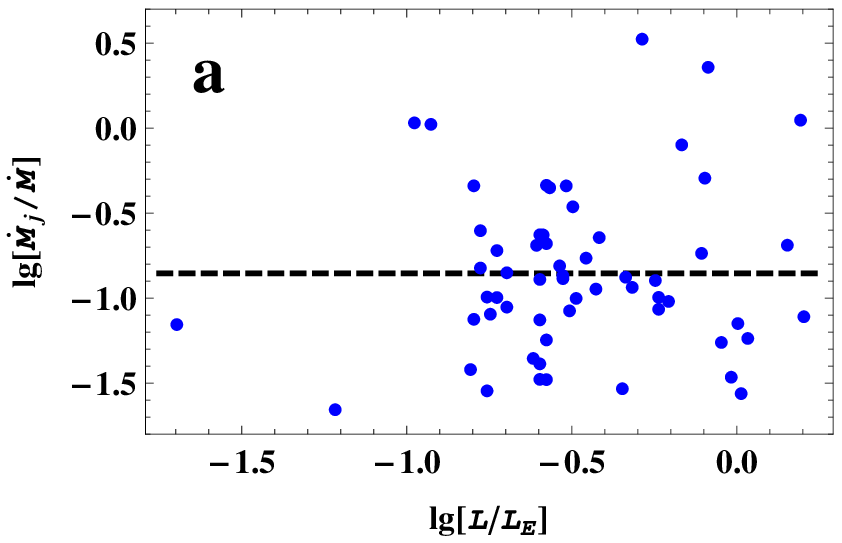}}
\centerline{\includegraphics[width=8cm]{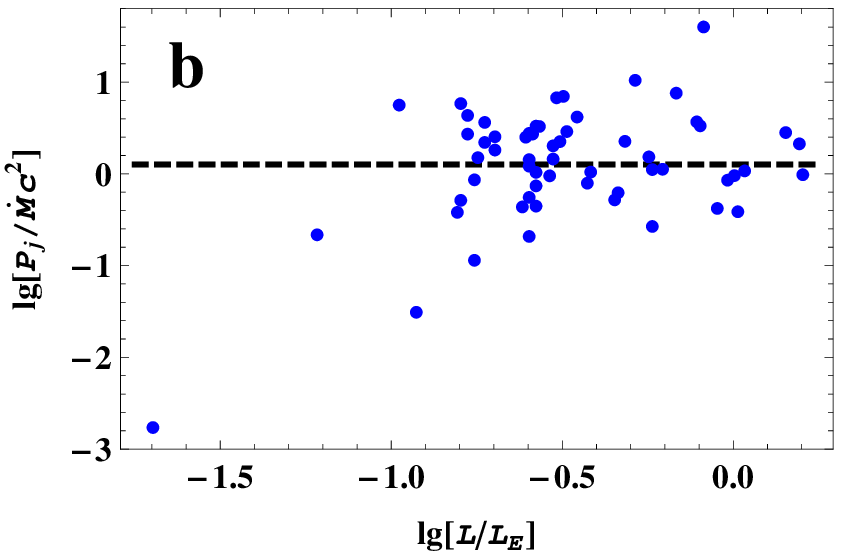}}
\caption{The dependence of the ratio of (a) the jet mass flow rate to $\dot M$ and (b) the total jet power to $\dot M c^2$ (for $\epsilon=0.2$) on the Eddington ratio for blazars. The dashed lines correspond to $\langle\dot M_{\rm j}/\dot M \rangle$ and $\langle P_{\rm j}/\dot M c^2\rangle$, respectively. 
}
\label{ratios}
\end{figure}

The constant total jet+counterjet mass-flow rate is $\dot M_{\rm j}= 2\upi \rho  c (\Theta_{\rm j}h)^2 \beta_{\rm j}\Gamma_{\rm j}$. Using equations (\ref{u_e}--\ref{open_sigma}) and (\ref{phi_jet}--\ref{phi_BH}), it can be written as
\begin{equation}
\dot M_{\rm j} = {(B h s)^2\beta_{\rm j}\over 2 c \Gamma_{\rm j}}\left(1-{\eta\beta\sigma\over 2}\right)={\phi_{\rm BH}^2\dot M r_{\rm g}^2\ell^2 a^2\beta_{\rm j}\over 16\upi^2 r_{\rm H}^2\Gamma_{\rm max}}.
\label{mdot}
\end{equation}
For the blazar sample of Z14 at $a=s=k=\beta=1$, we obtain $\langle \dot M_{\rm j}/\dot M\rangle\simeq 0.14(\epsilon/0.2)$, i.e., a relatively large fraction of the accretion flow is channelled into the jet. This further supports a modest value of the accretion radiative efficiency. Also, this strongly suggests that the jets are relatively heavy and their mass is dominated by baryons. This result depends only weakly on the unknown values of $\beta$. It may be $<1$, as a value $>1$ would imply an unlikely low average jet opening angle, see equation (\ref{B_beta}). Then, $\dot M_{\rm j}/\dot M\propto \phi_{\rm BH}^2 \propto B_\beta^2\propto \beta^{-4/9}$, which follows from equation (\ref{b1}), and which implies an even larger fractional $\dot M_{\rm j}$ for $\beta<1$. Fig.\ \ref{ratios}(a) shows the dependence of $\dot M_{\rm j}/\dot M$ on the Eddington ratio (for the H abundance of $X=0.7$). We see that most of the blazars have $L/L_{\rm E}$ in the 0.1--1 range, and there is no apparent trend of $\dot M_{\rm j}/\dot M$ seen within this interval. 

The jet+counterjet power, approximating $\Gamma_{\rm j}-1 \simeq \Gamma_{\rm j}$ in its rest-mass component, is $P=2\upi w c\beta_{\rm j}(\Theta_{\rm j}h\Gamma_{\rm j})^2$ (e.g., \citealt{levinson06}). We find the power in the magnetic field and the total power, respectively, as
\begin{eqnarray}
\lefteqn{
P_B={(B h s)^2\over 2}c\beta_{\rm j} \sigma=
{\phi_{\rm BH}^2\sigma\dot M c^2 r_{\rm g}^2\ell^2 a^2\beta_{\rm j}\over 16\upi^2 r_{\rm H}^2(1+\sigma)}\label{P_B}}\\
\lefteqn{
P_{\rm j}={(B h s)^2 \over 2}c\beta_{\rm j}(1+\sigma) = {\phi_{\rm BH}^2 \dot M c^2 r_{\rm g}^2 \ell^2 a^2\beta_{\rm j}\over 16\upi^2 r_{\rm H}^2}=\dot M_{\rm j}c^2 \Gamma_{\rm max},\label{P_tot}}
\end{eqnarray}
where we used equations (\ref{phi_jet}--\ref{mdot}). However, we note that some objects in both the blazar and radio galaxy samples of Z14 have $\beta_{\rm app}\ll 1$, and consequently $\Gamma_{\rm j}\sim 1$. While equation (\ref{P_B}) is then still valid, equation (\ref{P_tot}) no longer applies. We derive (based on the formulation of, e.g., \citealt{zdz14}),
\begin{equation}
P_{\rm j}=P_B\left(1+{\eta \beta\over 2\Gamma_{\rm j}}+{\Gamma_{\rm j}-1\over \sigma \Gamma_{\rm j}}\right).
\label{P_acc}
\end{equation}
We use equation (\ref{P_acc}) hereafter. 

For the sample of Z14 at $a=s=k=\beta=1$, we obtain $\langle P_{\rm j}/\dot M c^2\rangle\simeq 1.3(\epsilon/0.2)$ for blazars. This agrees well with the average from numerical simulations \citep{tnm11,mtb12}, as well as with the observational data in \citet{ghisellini14}. Fig.\ \ref{ratios}(b) shows the dependence of $P_{\rm j}/\dot M c^2$ on the Eddington ratio. In Figs.\ \ref{ratios}(a--b), we see a number of blazars with $\dot M_{\rm j}\ga \dot M$ and $P_{\rm j}/\dot M c^2\ga 3$, which may be due to the radiative efficiency lower than assumed, and/or inaccuracies in the estimated parameters, in particular $B$ and $L$. 

We point out that our result of $\langle P_{\rm j}/\dot M c^2\rangle\simeq 1.3(\epsilon/0.2)$ also strongly argues for the correctness of the identification of the measured $B$ with the radial average of the square of the field strength over the jet cross section, see equations (\ref{avb}), (\ref{bphiav}). If would instead identify the observed values with the surface magnetic field, $B'_{\phi,{\rm j}}$, the determined total jet power, which depends on the averaged field ($B$ in our notation), would be higher by a factor of $(1+\sigma)/(2\sigma)$, which, for our $\sigma\ll 1$ determined from the jet opening angles, would be $\gg \dot M c^2$, and thus clearly unphysical. 

We also note that in the case of minimization of the jet power with respect to the flux observed in the partially self-absorbed part for a fixed jet opening angle, the minimum jet power is achieved at $\sigma=(2 p+3)/10\simeq 0.5$ \citep{zdz14}. However, the opening angle is related to $\sigma$ in the present case and thus that result is not applicable.

\setlength{\tabcolsep}{4pt}
\begin{table*}
\begin{center}
\caption{The main parameters of the sample of the radio galaxies. We give the Eddington ratios based on Z14 (using their values of $M$), the values of the jet inclination, $i$, and the opening angle, $\Theta_{\rm j}$ (in radians), obtained by Z14 from literature (Savolainen and Zamaninasab, personal communication), and the corresponding values of $\Gamma_{\rm j}$ and $\vartheta_{\rm j}$ calculated using their values of $\beta_{\rm app}$. We also give our assumed values of $D_L$ and the 15 GHz flux based on literature. Then, we give the values found by us for magnetic field strength at 1 pc and the jet power based on $B_\beta$. 
}
\begin{tabular}{lcccccccccc}
\hline
Object &$D_L$\,[Mpc] &$L/L_{\rm E}$ & $\Gamma_{\rm j}$ & $i$ & $\Theta_{\rm j}$& $\vartheta_{\rm j}$ & $F_\nu$\,[Jy] & $B_\beta$\,[G] & $B_F$\,[G] & $P_{\rm j}$\,[erg\,s$^{-1}$]\\ 
\hline
0238--084 (NGC 1052)&19.4&  $7.6\times 10^{-6}$ &1.03 & 1.40 & 0.087&0.09	&2.4 &0.018 & $7.4\times 10^{-5}$ & $3.9\times 10^{41}$\\
0430+052 (3C 120)   &143 &  $2.4\times 10^{-1}$ &5.03 & 0.28 & 0.01 &0.05	&2.5 &0.15  & 0.0021 & $2.7\times 10^{45}$\\
0951+693 (M81)      &3.62&  $6.4\times 10^{-7}$ &2.09 & 0.25 & 0.009&0.02	&0.11&0.0079& 0.0029 & $4.0\times 10^{42}$\\
1219+285 (W Comae)  &465 &  $2.5\times 10^{-4}$ &7.16 & 0.44 & 0.09 &0.64	&0.57&0.097 & 0.0012 & $1.7\times 10^{45}$\\
1228+126 (M87)      &16.4&  $7.3\times 10^{-8}$ &1.16 & 0.24 & 0.261&0.30	&2.0 &0.014 & 0.016  & $4.2\times 10^{42}$\\
1322--427 (Cen A)   & 3.7&  $5.9\times 10^{-5}$ &1.12 & 1.13 & 0.017&0.02	&3.2 &0.020 & $2.5\times 10^{-4}$ & $2.7\times 10^{42}$\\
1845+797 (3C 390.3) &250 &  $4.1\times 10^{-3}$ &3.81 & 0.73 & 0.08 &0.30 	&0.31&0.11  & 0.012  & $1.3\times 10^{45}$\\
1957+405 (Cyg A)    &248 &  $9.4\times 10^{-4}$ &1.02 & 1.40 & 0.13 &0.13	&0.5 &0.088 & 0.014  & $1.1\times 10^{43}$\\
\hline
\end{tabular}
\end{center}
\label{rg}
\end{table*}

\section{Alternative magnetic-to-kinetic energy conversion}
\label{alternatives}

In the model adopted in Sections \ref{physics}--\ref{power}, the magnetic-to-kinetic energy flux conversion is assumed to result from the differential collimation of poloidal magnetic surfaces. This is the only currently known conversion mechanism in steady-state, axisymmetric and non-dissipative jets. (If there is a dissipative mechanism, the magnetic energy could be converted to thermal and kinetic energy.) However, the efficiency of this process is predicted to be very low for $\sigma < 1$ (e.g., \citealt{lyubarsky10}), and, therefore, it is not clear whether it can lead to $\sigma \ll 1$ in radio cores. Hence, it is likely that other mechanisms are involved in the conversion process, like reconnection driven by MHD instabilities (see \citealt{komissarov11} and references therein). The instabilities can develop when $\sigma$ drops to unity or even earlier if stimulated by high amplitude fluctuations of the jet power and direction, which are predicted in the MAD model (e.g., \citealt{mtb12}). In this case, $\sigma$ can reach values $\ll 1$ even prior to the blazar zone (i.e., the region where most of the radiation is produced), as indicated by blazar models with jet opening angles $\sim 1/\Gamma$ \citep*{nsb14,janiak15}. In such a scenario, collimation of a jet from $\Theta_{\rm j}\sim 1/\Gamma$ in the blazar zone down to $\Theta_{\rm j}\ll 1/\Gamma$ in the radio core can proceed following reconfinement of the flow by shocks triggered  by interaction of the cold, proton dominated, jets with the external medium. In this phase, the jet is supersonic, and, therefore, the reconfinement is not accompanied by the differential collimation of the poloidal magnetic surfaces and value of  $\sigma$ is not changing much.

Unfortunately, this conversion and collimation scenario does not offer us an opportunity to verify the MAD jet production mechanism by measuring the magnetic flux. This is because reconnection is expected to significantly affect the scaling of magnetic field with distance. Nevertheless, a strong argument in favour of the MAD scenario is provided by the energetics of quasar jets, often found to have the jet powers of the order of $\dot M c^2$ \citep{ghisellini10,ghisellini14}. Such jets require magnetic fluxes that are so large that can be supported on the BH only by the ram pressure of the accreting plasma.

If this alternative scenario is correct, the claimed agreement between the poloidal magnetic flux in radio cores and that threading the BH has to be accidental. However, we should note that this agreement is actually quite approximate, and it depends on the uncertain values of the radiative efficiency of the magnetically-dominated inner parts of the accretion flow and the uncertain values of the BH spin. 

Yet another alternative model has recently been proposed by \citet{nokhrina15}. While our results strongly indicate the jet composition is mostly electron-ion, they considered e$^\pm$ jets, but with only a small fraction, $\sim 0.01$, of the pairs being accelerated. This is likely to be difficult to reconcile with the observed lack of bulk-Compton spectral features, especially if the jet is accelerated up to $\Gamma_{\rm j} > 10$ already at distances $\la 10^4 r_{\rm g}$ \citep{sm00}.

\section{Radio galaxies}
\label{RG}

\begin{figure}
\centerline{\includegraphics[width=8cm]{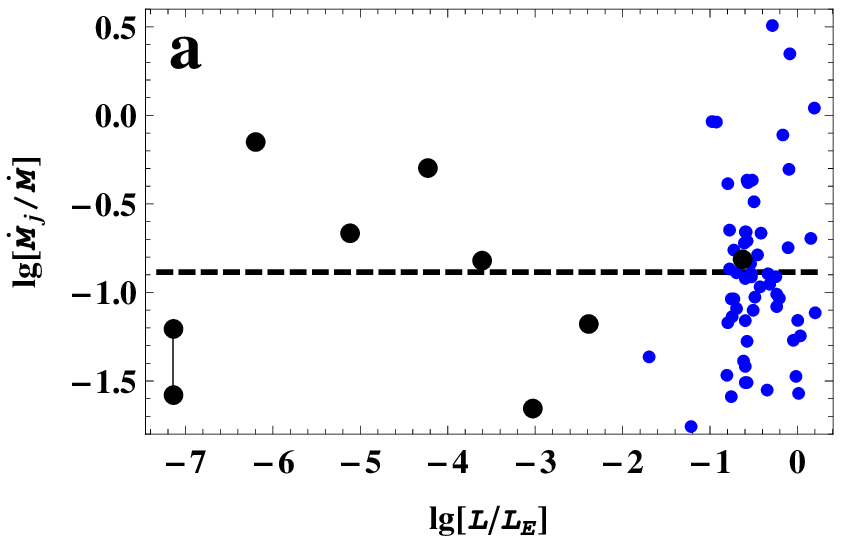}}
\centerline{{\hskip 0.3cm}\includegraphics[width=7.8cm]{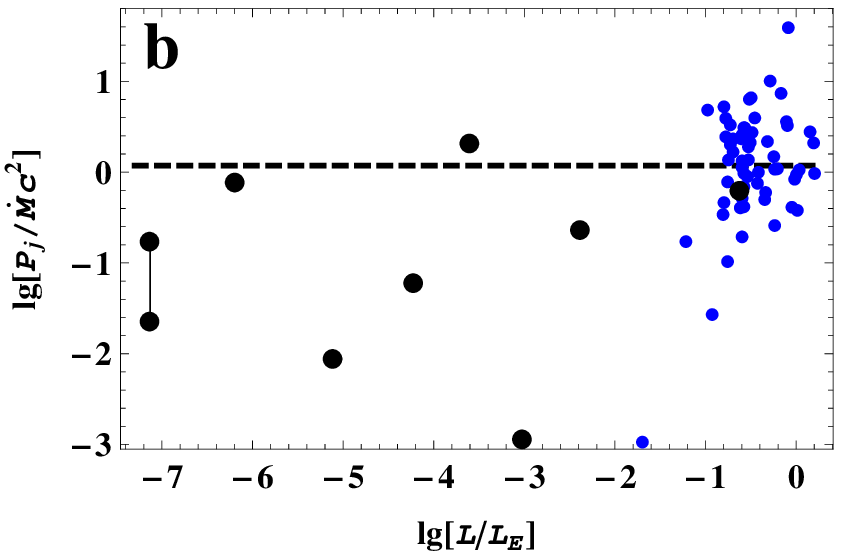}}
\caption{The dependence of the ratio of (a) the jet mass flow rate to $\dot M$ and (b) the total jet power to $\dot M c^2$ on the Eddington ratio for radio galaxies. The large black circles correspond to the radio galaxies using $B_\beta$ (at $\beta=1$). For M87, we show two pairs of values (connected by a line), with the upper one corresponding to $\beta_{\rm app}=6$, see Section \ref{RG}. For comparison, the small blue circles show the blazars (at $\beta=1$), and the dashed lines show their respective averages. The radiative efficiency of equation (\ref{epsilon}) has been used for all objects.
}
\label{ratios_rg}
\end{figure}

In calculating the inferred magnetic field at $h=1$ pc in radio galaxies, Z14 applied their equation (3) using their estimates of $i$ and $\Theta_{\rm j}$. Since Z14 did not provide those values, we list them for completeness in Table 1. Given the value of $i$, $\Gamma_{\rm j}(\beta_{\rm app})$ and $\delta$ are calculated, with $\beta_{\rm app}c$ given in Z14. Also, we note that the accretion luminosity of 3C 120 used by Z14 corresponds to X-rays only. For its broad-band spectrum, including UV, the value of $L\simeq 8\times 10^{44}$ erg s$^{-1}$ is found \citep{ogle05,kataoka11}, which we use in our calculations.

We have found the radio fluxes measured at the same time as the core shifts, and the distances (given that $z\ll 1$ for them) from literature, see Table 1. In Table 1 and Fig.\ \ref{B_ratio}, we see that all the radio galaxies have $B_F/B_\beta\la 1$, and $\langle B_F/B_\beta\rangle\simeq 0.05$. This is a strong systematic effect unlikely to be due to the measurement errors. At the face value, this indicates that those objects have magnetic fields different from blazars, and significantly below equipartition. In Table 1, we see that although $\vartheta_{\rm j}<1$ for all of the radio galaxies, there is a large range of its values. Still, the adopted values are not the cause of the obtained low values of $B_F/B_\beta$. We have tested this by assuming $\vartheta_{\rm j}=0.13$, in which case we have found $B_F/B_\beta< 1$ for most objects (except M81), and $\langle B_F/B_\beta \rangle \simeq 0.09$.

A possible reason for the above discrepancy and the difference with respect to blazars is the difference in viewing angles. Blazars are seen close to the jet axis, and the measured flux is strongly dominated by the fastest part of the jet, and, e.g., the flux from their radio lobes is completely negligible. On the other hand, radio galaxies are viewed from the side, and the relativistic beaming is minor. Thus, the measured radio flux may include slower parts of the jet and radio lobes. Thus, while our new method for measuring $B$ from the core shift works well for blazars, it is not reliable for radio galaxies unless we can clearly separate the part of the jet with the flat average spectrum (energy spectral index of $\simeq 0$), corresponding to the BK79 model. Thus, we no longer consider here the obtained values of $B_F$. We note that the magnetic field strengths in radio galaxies can also be estimated by using the angular sizes of the radio cores (which is outside the scope of this paper).

We find also that the radiative efficiency needs to be reduced for radio galaxies. Using the values of $B_\beta$ (at $\beta=1$) and $\epsilon=0.2$, we find $\dot M_{\rm j}/\dot M\gg 1$ for the 4 objects (out of 8) at the lowest $L/L_{\rm E}$. We thus take into account a decrease of $\epsilon$ at low accretion rates occuring in radiatively-inefficient accretion \citep{ny94}. We assume that the decrease of the radiative efficiency occurs at $\dot M\la \dot M_{\rm cr}=0.1 L_{\rm E}/c^2$,
\begin{equation}
\epsilon={\epsilon_0\over 1+0.1 L_{\rm E}/(\dot M c^2)}, \quad L=\epsilon \dot M c^2,
\label{epsilon}
\end{equation}
where $\epsilon_0=0.2$, as adopted for blazars in Section \ref{physics}. We then solve for $\dot M(L)$. The adopted value of $\dot M_{\rm cr}$ is relatively uncertain. In BH binaries, the above value corresponds to the typical transition from the soft to hard state (at $L \sim 0.02 L_{\rm E}$; \citealt{maccarone03}), but it may differ in AGNs. Since all objects in our samples have $L\la L_{\rm E}$, we do not include above the efficiency reduction of slim discs. 

Fig.\ \ref{ratios_rg}(a) shows the resulting values of $\dot M_{\rm j}/\dot M$. We see that taking into account the radiative inefficiency at low accretion rates results in the range of the obtained values similar to those for blazars. Fig.\ \ref{ratios_rg}(b) shows the obtained $P_{\rm j}/\dot M c^2$, and the values of $P_{\rm j}$ are given in Table 1. We see that the values are lower on average than those for blazars. The main cause of this difference appears to be the values of $\Gamma_{\rm j}$ being much lower in radio galaxies (see Table 1) than those in blazars. We note, however, that the observed low values of $\Gamma_{\rm j}$ in radio galaxies may result from their large viewing angles, at which any fast moving feature is relativistically de-boosted. Also, the values adopted in Z14 often correspond to a particular part of an observed jet. E.g., the value of $\Gamma_{\rm j}\simeq 1.16$ used for M87 corresponds to an inner jet, while the proper motion observed at larger distances implies $\Gamma_{\rm j}\sim 6$ \citep*{bsm99}. A larger $\Gamma_{\rm j}$ significantly increases the jet power. We show this in Fig.\ \ref{ratios_rg}, where the upper points for M87 correspond to $\beta_{\rm app}=6$ \citep{bsm99} and $\Theta_{\rm j}=0.01$ \citep{an12}, yielding $P_{\rm j}\simeq 3.2\times 10^{43}$ erg s$^{-1}$. Our results for the jet power can also be compared to the jet power determined by other methods, which, however, is outside the scope of this paper. 

We note that \citet{mg15} have also taken into account the radiative efficiency reduction at low accretion rates. They included only the jet power in the magnetic field [though their numerical coefficient for a single jet is lower by a factor of 0.63 from our value of $(1\,{\rm pc})^2 c/4$] and neglected the counterjet. Also, they neglected the power in particles, which, at their assumed $\sigma=1$, would introduce a further factor of 2 in the total jet power with respect to their equation (5). 

\section{Summary}
\label{summary}

We have derived an expression for the magnetic field using measurements of both the core shift and the total flux in the partially self-absorbed part of the spectrum (under the assumptions of the model of BK79), without assuming equipartition between the relativistic electrons and magnetic field. This expression is complementary to that using equipartition, and a comparison of them can be used to estimate the jet opening angle. For the latter formula, we have corrected a minor error in the power of $(1+z)$.

We have applied both formulae to the blazar sample of Z14. We have found that equipartition is approximately satisfied only if the jet opening angle $\Theta_j$ in the radio core region is close to the value found observationally, $\simeq (0.1$--$0.2)/\Gamma_{\rm j}$, in which case the magnetic field strengths are consistent with Z14. Larger values of $\Theta_j$ would imply very strong departures from equipartition, in particular $1/\Gamma_{\rm j}$ yields $\beta\sim 10^{-9}$.

We have then studied the magnetic flux in the jet in the framework of the model of \citet{bz77} following the method of Z14, but accounting for the obtained low value of the jet opening angle, implying the magnetization parameter of $\sigma\ll 1$. Also, we expressed the magnetic flux in terms of the transverse average of the toroidal field strength, which appears to correspond to the observed values more directly than the surface strength used before. We have confirmed that the average jet magnetic flux is compatible with both the magnetically arrested accretion and $\sigma\ll 1$. This helps to resolve some of the problems with modelling of blazars for the strong magnetic fields measured by the core shift method pointed out by \citet{nsb14}.

We have calculated the jet mass-flow corresponding to the above model, and found it consists of a substantial fraction of the mass accretion rate, $\dot M_{\rm j}\simeq 0.14\dot M$ for our chosen parameters. For those, the jet power is $P_{\rm j}\simeq 1.3\dot M c^2$, consistent with the results of numerical simulations of MADs. 

Finally, we have found that at least some of the radio galaxies in the sample of Z14 may also accrete via the MADs. However, the accretion is required to be radiatively inefficient, in agreement with standard models. 

\section*{Acknowledgments}

We thank Alexander Pushkarev for help with using the radio data for blazars, Tuomas Savolainen for his kind help with the parameters of the sample of the radio galaxies of Z14, and Arieh K{\"o}nigl for valuable comments on this work. This research has made use of data from the MOJAVE database that is maintained by the MOJAVE team, and it has been supported in part by the Polish NCN grants 2012/04/M/ST9/00780, 2013/10/M/ST9/00729 and DEC-2011/01/B/ST9/04845. AT was supported by NASA through Einstein Postdoctoral Fellowship grant number PF3-140115 awarded by the Chandra X-ray Center, which is operated by the Smithsonian Astrophysical Observatory for NASA under contract NAS8-03060.

\label{lastpage}


\begin{thebibliography}{}

\bibitem[\protect\citeauthoryear{Asada \& Nakamura}{2012}]{an12} 
Asada K., Nakamura M., 2012, ApJ, 745, L28 

\bibitem[\protect\citeauthoryear{Biretta, Sparks \& Macchetto}{Biretta et al.}{1999}]{bsm99} 
Biretta J.~A., Sparks W.~B., Macchetto F., 1999, ApJ, 520, 621 

\bibitem[\protect\citeauthoryear{Blandford \& K{\"o}nigl}{1979}]{bk79} 
Blandford R.~D., K{\"o}nigl A., 1979, ApJ, 232, 34 (BK79)

\bibitem[\protect\citeauthoryear{Blandford \& Znajek}{1977}]{bz77} 
Blandford R.~D., Znajek R.~L., 1977, MNRAS, 179, 433 

\bibitem[\protect\citeauthoryear{Clausen-Brown et al.}{2013}]{clausen13} 
Clausen-Brown E., Savolainen T., Pushkarev A.~B., Kovalev Y.~Y., Zensus J.~A., 2013, A\&A, 558, A144 

\bibitem[\protect\citeauthoryear{Ghisellini et al.}{2010}]{ghisellini10} 
Ghisellini G., Tavecchio F., Foschini L., Ghirlanda G., Maraschi L., Celotti A., 2010, MNRAS, 402, 497 

\bibitem[\protect\citeauthoryear{Ghisellini et al.}{2014}]{ghisellini14} 
Ghisellini G., Tavecchio F., Maraschi L., Celotti A., Sbarrato T., 2014, Nature, 515, 376

\bibitem[\protect\citeauthoryear{Hirotani}{2005}]{hirotani05} 
Hirotani K., 2005, ApJ, 619, 73 

\bibitem[\protect\citeauthoryear{Janiak, Sikora \& Moderski}{Janiak et al.}{2015}]{janiak15} 
Janiak M., Sikora M., Moderski R., 2015, MNRAS, 449, 431 

\bibitem[\protect\citeauthoryear{Jorstad et al.}{2005}]{jorstad05} 
Jorstad S.~G., et al., 2005, AJ, 130, 1418 

\bibitem[\protect\citeauthoryear{Kataoka et al.}{2011}]{kataoka11} 
Kataoka J., et al., 2011, ApJ, 740, 29 

\bibitem[\protect\citeauthoryear{Komissarov}{2011}]{komissarov11} 
Komissarov S.~S., 2011, Mem.\ Soc.\ Astron.\ Ital., 82, 95 

\bibitem[\protect\citeauthoryear{Komissarov et al.}{2009}]{komissarov09} 
Komissarov S.~S., Vlahakis N., K{\"o}nigl A., Barkov M.~V., 2009, MNRAS, 394, 1182

\bibitem[\protect\citeauthoryear{K{\"o}nigl}{1981}]{konigl81} 
K{\"o}nigl A., 1981, ApJ, 243, 700 

\bibitem[\protect\citeauthoryear{Kovalev et al.}{2005}]{kovalev05} 
Kovalev Y.~Y., et al., 2005, AJ, 130, 2473 

\bibitem[\protect\citeauthoryear{Leahy}{1991}]{leahy91} 
Leahy J.~P., 1991, in P. A. Hughes, ed., Beams and Jets in Astrophysics. Cambridge Univ.\ Press, Cambridge, p.\ 100 

\bibitem[\protect\citeauthoryear{Levinson}{2006}]{levinson06} 
Levinson A., 2006, Int.\ J. Mod.\ Phys., 21, 6015 

\bibitem[\protect\citeauthoryear{Lister et al.}{2009}]{lister09} 
Lister M.~L., et al., 2009, AJ, 137, 3718 

\bibitem[\protect\citeauthoryear{Lobanov}{1998}]{lobanov98}
Lobanov A. P., 1998, A\&A, 330, 79 

\bibitem[\protect\citeauthoryear{Lyubarsky}{2010}]{lyubarsky10} 
Lyubarsky Y.~E., 2010, MNRAS, 402, 353 

\bibitem[\protect\citeauthoryear{Maccarone}{2003}]{maccarone03} 
Maccarone T.~J., 2003, A\&A, 409, 697 

\bibitem[\protect\citeauthoryear{McKinney, Tchekhovskoy \& Blandford}{McKinney et al.}{2012}]{mtb12} 
McKinney J.~C., Tchekhovskoy A., Blandford R.~D., 2012, MNRAS, 423, 3083 

\bibitem[\protect\citeauthoryear{Miller-Jones, Fender \& Nakar}{Miller-Jones et al.}{2006}]{millerjones06} 
Miller-Jones J.~C.~A., Fender R.~P., Nakar E., 2006, MNRAS, 367, 1432 

\bibitem[\protect\citeauthoryear{Mocz \& Guo}{2015}]{mg15} 
Mocz P., Guo X., 2015, MNRAS, 447, 1498 

\bibitem[\protect\citeauthoryear{Nalewajko, Sikora \& Begelman}{Nalewajko et al.}{2014}]{nsb14} 
Nalewajko K., Sikora M., Begelman M. C., 2014, ApJ, 796, L5

\bibitem[\protect\citeauthoryear{Narayan \& Yi}{1994}]{ny94} 
Narayan R., Yi I., 1994, ApJ, 428, L13 

\bibitem[\protect\citeauthoryear{Narayan, Igumenshchev \& Abramowicz}{Narayan et al.}{2003}]{nia03} 
Narayan R., Igumenshchev I.~V., Abramowicz M.~A., 2003, PASJ, 55, L69 

\bibitem[\protect\citeauthoryear{Nokhrina et al.}{2015}]{nokhrina15} 
Nokhrina E.~E., Beskin V.~S., Kovalev Y.~Y., Zheltoukhov A.~A., 2015, MNRAS, 447, 2726 

\bibitem[\protect\citeauthoryear{Ogle et al.}{2005}]{ogle05} 
Ogle P.~M., Davis S.~W., Antonucci R.~R.~J., Colbert J.~W., Malkan M.~A., Page M.~J., Sasseen T.~P., Tornikoski M., 2005, ApJ, 618, 139 

\bibitem[\protect\citeauthoryear{O'Sullivan \& Gabuzda}{2009}]{og09} 
O'Sullivan S.~P., Gabuzda D.~C., 2009, MNRAS, 400, 26 

\bibitem[\protect\citeauthoryear{Pushkarev et al.}{2009}]{pushkarev09} 
Pushkarev A.~B., Kovalev Y.~Y., Lister M.~L., Savolainen T., 2009, A\&A, 507, L33 

\bibitem[\protect\citeauthoryear{Pushkarev et al.}{2012}]{pushkarev12} 
Pushkarev A.~B., Hovatta T., Kovalev Y.~Y., Lister M.~L., Lobanov A.~P., Savolainen T., Zensus J.~A., 2012, A\&A, 545, A113 

\bibitem[\protect\citeauthoryear{Sikora \& Madejski}{2000}]{sm00} 
Sikora M., Madejski G., 2000, ApJ, 534, 109 

\bibitem[\protect\citeauthoryear{Slish}{1963}]{slish63} 
Slish V.~I., 1963, Nature, 199, 682 

\bibitem[\protect\citeauthoryear{Tchekhovskoy}{2015}]{t15} 
Tchekhovskoy A., 2015, in I. Contopoulos et al, eds., The Formation and Disruption of Black Hole Jets. ASSL, 414, 45

\bibitem[\protect\citeauthoryear{Tchekhovskoy \& McKinney}{2012}]{tm12} 
Tchekhovskoy A., McKinney J.~C., 2012, MNRAS, 423, L55 

\bibitem[\protect\citeauthoryear{Tchekhovskoy, McKinney \& Narayan}{Tchekhovskoy et al.}{2009}]{tmn09} 
Tchekhovskoy A., McKinney J.~C., Narayan R., 2009, ApJ, 699, 1789 

\bibitem[\protect\citeauthoryear{Tchekhovskoy, Narayan \& McKinney}{Tchekhovskoy et al.}{2011}]{tnm11} 
Tchekhovskoy A., Narayan R., McKinney J.~C., 2011, MNRAS, 418, L79 

\bibitem[\protect\citeauthoryear{Williams}{1963}]{williams63} 
Williams P.~J.~S., 1963, Nature, 200, 56 

\bibitem[\protect\citeauthoryear{Zamaninasab et al.}{2014}]{zamaninasab14} 
Zamaninasab M., Clausen-Brown E., Savolainen T., Tchekhovskoy A., 2014, Nature, 510, 126 (Z14)

\bibitem[\protect\citeauthoryear{Zdziarski}{2014}]{zdz14} 
Zdziarski A.~A., 2014, MNRAS, 445, 1321

\bibitem[\protect\citeauthoryear{Zdziarski, Lubi{\'n}ski \& Sikora}{Zdziarski et al.}{2012}]{zls12} 
Zdziarski A.~A., Lubi{\'n}ski P., Sikora M., 2012, MNRAS, 423, 663 (ZLS12)

\end{thebibliography}
\end{document}